\documentclass[12pt]{article}
\usepackage{euscript,amsmath, amssymb, amsfonts}
\usepackage{color}

\newcommand{\R}{{\mathbb R}}
\newcommand{ \nn}{\nonumber \\}
\newcommand{ \bee}{\begin{eqnarray}}
\newcommand{ \eee}{\end{eqnarray}}

\newcommand{\p}{{\hbar^2}}

\newcommand{\oC}{{\mathbb C}}
\newcommand{\bbreve}[1]{{\mathbf #1}}

\newcommand{\di}{{ d }}

\newcommand{\eE}{\EuScript E}

\newcommand{\G}{{\mathbb G}}
\newcommand{\Z}{{\mathbb Z}}
\newcommand{\K}{{\mathbb K}}
\newcommand{\be}{\begin{equation}}
\newcommand{\ee}{\end{equation}}

\newcounter{theorem}
\newcommand{\theorem}{\par\refstepcounter{theorem}
           {\bf Theorem \arabic{section}.\arabic{theorem}. }}

\makeatletter \@addtoreset{theorem}{section}

\newcounter{lemma}

\makeatletter \@addtoreset{lemma}{section}

\newcounter{proposition}

\makeatletter \@addtoreset{proposition}{section}

\makeatletter \@addtoreset{equation}{section}

\makeatother

\newcounter{appen}

\newcounter{subappen}

\makeatletter \@addtoreset{subappen}{appen}

\newcounter{subsubappen}

\makeatletter \@addtoreset{subsubappen}{subappen}

\font\frtnfr=eufm10   scaled\magstep1
\font\twlfr=eufm10
\font\tenfr=eufm10

\newfam\frfam
\textfont\frfam=\frtnfr
\scriptfont\frfam=\twlfr
\scriptscriptfont\frfam=\tenfr

\font\frtnopen=msbm10  scaled\magstep2
\font\twlopen=msbm10
\font\tenopen=msbm10

\newfam\openfam
\textfont\openfam=\frtnopen
\scriptfont\openfam=\twlopen
\scriptscriptfont\openfam=\tenopen

\font\frtnsf = cmss12 scaled\magstep1
\font\twlsf = cmss10
\font\tensf = cmss9

\newfam\Scfam
\textfont\Scfam = \frtnsf
\scriptfont\Scfam = \twlsf
\scriptscriptfont\Scfam = \tensf

\begin{document}

\sloppy \title
 {
     Deformations of the antibracket.
 }
\author
 {
 S.E.Konstein\thanks{E-mail: konstein@lpi.ru}\ \ and
 I.V.Tyutin\thanks{E-mail: tyutin@lpi.ru}
 \thanks{
               This work was supported
               by the RFBR (grants No.~05-01-00996
               (I.T.) and No.~05-02-17217 (S.K.)),
               and by the grant LSS-1578.2003.2.
 } \\
               {\sf \small I.E.Tamm Department of
               Theoretical Physics,} \\ {\sf \small P. N. Lebedev Physical
               Institute,} \\ {\sf \small 119991, Leninsky Prospect 53,
               Moscow, Russia.} }
\date {}

\maketitle

\begin{abstract}
{ \footnotesize We consider antiPoisson superalgebras realized on the smooth
Grassmann-valued functions with compact supports in $\R^n$ and with the grading
inverse
to Grassmanian parity.
The deformations of these superalgebras and their central extensions
are found.
}
\end{abstract}


\section{Introduction}

The goal of present work is to find the deformations
of antiPoisson superalgebra realized on
the smooth Grassmann-valued functions with
compact supports in $\R^n$. The results are based on the results of
\cite{cohom} where the lower cohomologies of this
antiPoisson superalgebra are found. Particularly, it is shown here
that the nontrivial deformations of antiPoisson superalgebras do exist, and that
the central extensions of antiPoisson superalgebra (which exist for $n=1$) have
no nontrivial deformations.
The deformations of Poisson superalgebra
realized on the space of vector fields with polynomial coefficients and
with antibracket were obtained in
\cite{Leites}.

Let $\K$ be either $\R$ or $\oC$.
We denote by $\EuScript D(\R^n)$ the space
of smooth $\K$-valued functions with compact supports on $\R^n$.
This space is
endowed with its standard topology: by definition,
a sequence $\varphi_k\in
\EuScript D(\R^n)$ converges to $\varphi\in \EuScript D(\R^n)$
if the
supports of all $\varphi_k$ are contained in a fixed compact set,
and
$\partial^\lambda\varphi_k$ converge uniformly to $\partial^\lambda\varphi$
for every multi-index $\lambda$.
We set
$
 \mathbf D_{n}= \EuScript
D(\R^{n})\otimes \G^{n}
,
$
where $\G^{n}$ is the Grassmann
algebra with $n$ generators. The generators of
the Grassmann algebra (resp., the coordinates of the space $\R^{n}$) are
denoted by $\xi^\alpha$, $\alpha=1,\ldots,n$ (resp., $x^i$, $i=1,\ldots,
n$).
We shall also use collective variables $z^A$ which are equal to $x^A$
for $A=1,\ldots,n$ and are equal to $\xi^{A-n}$ for
$A=n+1,\,\ldots,\,2n$.

The space $\mathbf D_{n}$ possesses a natural grading
which is determined by that of the Grassmann algebra.
The Grassmann parity (below -- $\varepsilon$-parity) of an
element $f$ of these spaces is denoted by $\varepsilon(f)$.

The space $\mathbf D_{n}$ possesses also another $\Z_2$-grading
$\epsilon$ ($\epsilon$-parity),
 which is inverse to $\varepsilon$-parity: $\epsilon=\varepsilon+1$.

We set
$\varepsilon_A=0$, $\epsilon_A=1$ for $A=1,\ldots, n$
and $\varepsilon_A=1$, $\epsilon_A=0$ for
$A=n+1,\,\ldots, \,2n$.

It is well known, that the bracket
\bee\label{Sch}
[f,g]=\sum_{i=1}^n \left(
f \frac{\overleftarrow{\partial}}{\partial x^i} \frac\partial {\partial \xi^i} g
- f \frac{\overleftarrow{\partial}}{\partial \xi^i}\frac \partial{\partial x^i} g
\right),
\eee
which we will call "antibracket",
defines the structure of Lie superalgebra on the superspace
$\mathbf D_{n}$
with the $\epsilon$-parity.

Indeed, $[f,g]=-(-1)^{\epsilon(f)\epsilon(g)}[g,f]$,
$\epsilon([f,g])=\epsilon(f)+\epsilon(g)$,
and Jacobi identity is satisfied:
\be\label{3.0a}
(-1)^{\epsilon(f)\epsilon(h)}[f,[g,h]]
+(-1)^{\epsilon(g)\epsilon(f)}[g,[h,f]]
+(-1)^{\epsilon(h)\epsilon(g)}[h,[f,g]]=0
,\quad f,g,h\in \mathbf D_{n}.
\ee

Here this Lie superalgebra is called antiPoisson
superalgebra.\footnote{
We will consider usual multiplication of the elements of considered antiPoisson
superalgebras with commutation relations $fg=(-1)^{\varepsilon(f)\varepsilon(g)}gf$
as well, and the variables $x^i$ will be called even variables and the variables
$\xi^i$ will be called odd variables.}

The integral on $\mathbf D_{n}$ is defined by the relation $ \int \di
z\, f(z)= \int_{\R^{n}}\di x\int \di\xi\, f(z), $ where the integral on the
Grassmann algebra is normed by the condition $\int \di\xi\,
\xi^1\ldots\xi^{n}=1$. As a rule, the value $m(f)$
of a functional $m\in \mathbf D^{\prime}_{n}$ on a test function $f\in
\mathbf D_{n}$ will be written in the integral form:  $ m(f)
=\int \di z\, m(z) f(z).  $

The second cohomology
space\footnote
{The definition of the differentials $d_p^{\mathrm {ad}}$ see in \cite{cohom}}
 $H^2_{\mathrm{ad}}$ in the adjoint representation is closely related to
the problem of finding formal deformations of the Lie bracket
$[\cdot,\cdot]$ of the form $ [f,g]_*=[f,g]+\hbar [f,g]_1+\ldots \equiv C(f,g)=
C_i(f,g)\hbar^i. $
The condition that $C_1(f,g)$ is a 2-cocycle is equivalent to the
Jacobi identity for $[\cdot,\cdot]_*$ modulo the $\hbar^2$-order terms.

Formal deformations $C^1$ and $C^2$ are called equivalent if there is a
continuous $\K[[\p]]$-linear similarity operator $T: \mathbf D_{n}[[\p]]\to
\mathbf D_{n}[[\p]]$ such that
$TC^1(f,g)=C^2(T f,Tg)$, $f,g\in \mathbf D_n[[\p]]$ and $T=id +\hbar T_1$.

The following theorem is proved in \cite{cohom}

\theorem\label{thad}{}
{\it
Let bilinear mappings $m$ and
$n$ from
$(\mathbf D_n)^2$ to $\mathbf D_{n}$
be defined by the
relations
\bee
m(z|f,g)&=&(-1)^{\varepsilon (f)}\{\Delta f(z)\}\eE_z g(z)+
\{\eE_z f(z)\}\Delta g(z)
\\
n(z|f,g)&=&(-1)^{\varepsilon (f)}\{(1-N_{\xi })f(z)\}(1-N_{\xi })g(z),
\eee
where
\bee
N_\xi=\xi^i\partial_{\xi^i},\\
\eE_z = 1-\frac 1 2 z^A \frac \partial {\partial z^A},\\
\Delta=
\frac \partial {\partial x^i}\frac \partial {\partial \xi^i}.
\label{Delta}
\eee
Then
$H^2_{\mathrm {ad}}\simeq \K^2$ and the cochains
$m(z|f,g)$ and $n(z|f,g)$
are
independent nontrivial cocycles.
}

This theorem allows us to prove here the following theorem, stating the
general form of the deformation of antiPoisson superalgebra:

\theorem\label{thdef}{}
{\it The deformation of antiPoisson superalgebra with even parameter $\hbar$
has the form up to similarity transformation
\begin{equation}\label{def}
[f(z),\, g(z)]_*\!=\![f(z),g(z)] + (-1)^{\varepsilon(f)}\{\frac{\hbar
c}{1+\hbar cN_z/2}\Delta f(z)\} \eE_zg(z)
+
\{\eE_z f(z)\}
\frac{\hbar c}{1+\hbar cN_z/2}\Delta g(z),
\end{equation}
where $N_z=z^A \frac {\partial}{\partial z^A}$, and $c$ is an arbitrary
formal series
in $\hbar$ with coefficients in $\K$.}

It follows from the next theorem proved in \cite{cohom}
that if $n=1$ then antiPoisson superalgebra has a central extension:
\theorem\label{thtr}{}
{\it \begin{enumerate}
\item
Let $n\geq 2$. Then $H^2_{\mathrm{tr}}\simeq 0$.
\item
Let $n=1$. Then $H^2_{\mathrm{tr}}\simeq \K^2$, and the cochains
\bee
&&\mu _{1}(f,g)=\int dz(-1)^{\varepsilon (f)}\{\partial_{x}^{3}\partial _{\xi
}f(z)\}g(z),\nn
&&\mu _{2}(f,g) =\int dz(-1)^{\varepsilon (f)}\{\partial
_{x}^{2}f(z)\}g(z)\label{mu}
\eee
are
independent nontrivial cocycles.
\end{enumerate}
}

It is proved below that
this central extension has no deformation with even parameter $\hbar$.

\section{Deformations of the antibracket}

Consider general form of deformation, $[f,g]_{\ast }(z)$, of the antibracket
$[f,g](z)$.

We will suppose that:

1)
\begin{eqnarray}
[f,g]_*(z) &\equiv &C(z|f,g)=\sum_{k=0}\hbar^{k}C_{k}(z|f,g), \label{decc}\\
 C_{0}(z|f,g)&=& \lbrack f,g\rbrack (z);
\end{eqnarray}

2)
\begin{equation*}
C_{k}(z|f,g)=\int dvdu(-1)^{\varepsilon (f)}C_{k}(z|u,v)f(u)g(v);
\end{equation*}

3)
\begin{equation*}
\varepsilon (C(z|f,g))=\varepsilon (C_{k}(z|f,g))=\varepsilon
(f)+\varepsilon (g)+1;
\end{equation*}

4)
\begin{eqnarray*}
C(z|g,f) &=&-(-1)^{\epsilon (f)\epsilon (g)}C(z|f,g), \\
C_{k}(z|v,u)&=&(-1)^{n}C_{k}(z|u^*,v^*),
\mbox{ where if $u=(x,\xi)$ then $u^*=(x,-\xi)$;}
\end{eqnarray*}

5) $C(z|g,f)\in \mathbf D_n$, $C_{k}(z|f,g)\in \mathbf D_n$
for all $f,g \in \mathbf D_n$;

6) $[f,g]_{\ast }(z)$ satisfies the Jacobi identity
\begin{equation}
(-1)^{\epsilon (f)\epsilon (h)}[[f,g]_{\ast },h]_{\ast }+
\mathrm{cycle}(f,g,h)=0,\;\forall f,g,h,  \label{1.1}
\end{equation}
or
\begin{equation}
(-1)^{\epsilon (f)\epsilon (h)}C(z|C(|f,g),h)+\mathrm{cycle}
(f,g,h)=0.  \label{1.1a}
\end{equation}

Note that if a form $C(z|f,g)$ satisfies the Jacobi identity then the form
$C_{T}(z|f,g)$,
\begin{equation*}
C_{T}(z|f,g)=T^{-1}C(z|Tf,Tg),
\end{equation*}
satisfies the Jacobi identity too. Here $T$: $f(z)\rightarrow T(z|f)$ is
invertible continuous map $\mathbf D_n\rightarrow \mathbf D_n$.

\subsection{$\hbar $-order}

We have from Eq, (\ref{1.1a})
\begin{eqnarray*}
&&(-1)^{\epsilon (f)\epsilon
(h)}[C_{1}(z|f,g),h(z)]+(-1)^{\epsilon (f)\epsilon
(h)}C_{1}(z|[f,g)],h)+ \\
&&\,+\mathrm{cycle}(f,g,h)=-(-1)^{\epsilon (f)\epsilon
(h)}d_{2}^{\mathrm{ad}}C_{1}(z|f,g,h)=0,\;\varepsilon
_{C_{1}}=1,
\end{eqnarray*}
so theorem \ref{thad} gives
\begin{equation*}
C_{1}(z|f,g)=c_{1}m(z|f,g)+d_{1}^{\mathrm{ad}}M_{1D}(z|f,g)+\alpha_{1}n(z|f,g),
\end{equation*}
where $m$ and $n$ are defined in theorem \ref{thad}.

Because we consider the deformations with even parameters only,
and because $\varepsilon(\alpha_1)=1$, we set $\alpha_1=0$. Analogously,
$\alpha_i=0$, where $\alpha_i$ are the coefficients before $m_2$ in
higher orders on $\hbar$, {\it i.e.} we don't take in account the
cocycle $n$.

The Jacobiator has the form
\begin{eqnarray*}
&&J_{m}(z|f,g,h)\equiv (-1)^{\epsilon (f)\epsilon
(h)}m(z|m(|f,g),h)+\mathrm{cycle}(f,g,h)= \\
&&=(-1)^{\varepsilon (f)+\varepsilon (g)+\varepsilon (h)+\varepsilon
(f)\varepsilon (h)}\{[\eE_zf(z),\Delta g(z)]\eE_zh(z)-(-1)^{\varepsilon
(f)}[\Delta f(z),\eE_zg(z)]\eE_zh(z)\}+ \\
&&+\mathrm{cycle}(f,g,h).
\end{eqnarray*}
For the functions of the form $f(z)= e^{zp}$, $g(z)=
e^{zq}$, $h(z)= e^{zr}$ in some vicinity of $x$, we have
\begin{eqnarray*}
&&J_m(z|f,g,h)\;\rightarrow \;J_{m}(z|p,q,r)= \\
&&\,=-\frac{1}{4}[\left\langle p,p\right\rangle (1-zq)+\left\langle
q,q\right\rangle (1-zp)](1-\frac{1}{2}zr)\left\langle p,q\right\rangle +
\mathrm{cycle}(p,q,r),
\end{eqnarray*}
where
$$
\langle p,\,q \rangle = \sum_{i=1}^n \left(p^i q^{n+i}+ q^i p^{n+i}\right).
$$

\subsection{$\hbar ^{2}$-order}

We have from Eq, (\ref{1.1a})
after $T$-transformation, $T=1+\hbar M_{1D}+O(\hbar^2)$,
\begin{eqnarray*}
&&(-1)^{\epsilon (f)\epsilon
(h)}[C_{2}(z|f,g),h(z)]+(-1)^{\epsilon (f)\epsilon
(h)}C_{2}(z|[f,g]),h)+ \\
&&\,+\mathrm{cycle}(f,g,h)=-(-1)^{\epsilon (f)\epsilon
(h)}d_{2}^{\mathrm{ad}}C_{2}(z|f,g,h)= \\
&&\,=-c_{1}^{2}J_{m}(f,g,h),\;\varepsilon _{C_{2}}=1.
\end{eqnarray*}
Because $J_{m_{1}}(z|f,g,h)$ equals to zero out of any diagonal, we have
\begin{eqnarray*}
&&C_{2}(z|f,g)=C_{2\mathrm{loc}}(z|f,g)+d_{1}^{\mathrm{ad}}M_{2D}(z|f,g)
,
\end{eqnarray*}
where $C_{2\mathrm{loc}}(z|f,g)$ has the form
\begin{eqnarray*}
&&C_{2\mathrm{loc}}(z|f,g)= \\
&&\,=\sum_{a,b=0}^{N}(-1)^{\varepsilon (f)(|\varepsilon
_{B}|_{1,b}+1)}m_{2}^{(A)_{a}|(B)_{b}}(z)[(\partial
_{A}^{z})^{a}f(z)](\partial _{B}^{z})^{b}g(z), \\
&&m_{2}^{(B)_{q}|(A)_{p}}=(-1)^{|\varepsilon _{A}|_{1,p}|\varepsilon
_{B}|_{1,q}}m_{2}^{(A)_{p}|(B)_{q}},\;\varepsilon
(m_{2}^{(A)_{p}|(B)_{q}}(\partial _{A}^{z})^{a}(\partial _{B}^{z})^{b})=1,
\end{eqnarray*}
\begin{equation*}
(-1)^{\epsilon (f)\epsilon (h)}d_{2}^{\mathrm{ad}
}C_{2\mathrm{loc}}(z|f,g,h)=c_{1}^{2}J_{m_{1}}(f,g,h).
\end{equation*}
Take the functions in the form $f(z)= e^{zp}$, $g(z)=
e^{zq}$, $h(z)= e^{zr}$ in some vicinity of $x$. Then we have
\begin{eqnarray}
&&\Phi (z,p,q,r)\left\langle p,q\right\rangle -[F(z,p,q),zr]+\mathrm{cycle}
(p,q,r)=  \notag \\
&&\,=\frac{c_{1}^{2}}{4}[\left\langle p,p\right\rangle (1-zq)+\left\langle
q,q\right\rangle (1-zp)](1-\frac{1}{2}zr)\left\langle p,q\right\rangle +
\mathrm{cycle}(p,q,r),  \label{3.1}
\eee
where
\bee
&&\Phi (z,p,q,r)=\Phi (z,q,p,r)=F(z,p+q,r)-F(z,p,r)-F(z,q,r),  \notag \\
&&F(z,p,q)=F(z,q,p)=
\sum_{a,b=0}^{N}m_{2}^{(A)_{a}|(B)_{b}}(z)(p_{A})^{a}(q_{B})^{b}.  \notag
\end{eqnarray}

Represent $F(z,p,q)$ in the form
\begin{eqnarray*}
&&F(z,p,q)=-\frac{c_{1}^{2}}{4}[\left\langle p,p\right\rangle
(zp)+\left\langle q,q\right\rangle (zq)]+ \\
&&+\frac{c_{1}^{2}}{8}(\left\langle p,p\right\rangle +\left\langle
q,q\right\rangle )(zp)(zq)+F^{\prime }(z,p,q), \\
&&F^{\prime }(z,p,q)=\sum_{a,b=0}^{N}m_{2}^{\prime
(A)_{a}|(B)_{b}}(z)(p_{A})^{a}(q_{B})^{b}.
\end{eqnarray*}
Then we obtain
\begin{equation*}
\Phi ^{\prime }(z,p,q,r)\left\langle p,q\right\rangle -[F^{\prime
}(z,p,q),zr]+\mathrm{cycle}(p,q,r)=0.
\end{equation*}
Using the results of subsection 5.2 in \cite{cohom}, we
find
\begin{eqnarray*}
&&C_{2|\mathrm{loc}}(z|f,g)=c_{2}m(z|f,g)-\frac{c_{1}^{2}}{2}
m_{2}(z|f,g)+d_{1}^{\mathrm{ad}}M_{1D}(z|f,g),
\end{eqnarray*}
where
\bee
m_{2}(z|f,g)&=&
(-1)^{\varepsilon (f)}\{N_{z}\Delta f(z)\}\eE_zg(z)+\{
\eE_zf(z)\}N_{z}\Delta g(z).\notag
\eee

The conjecture arises that deformation depends on Euler operator $N$
and nilpotent operator $\Delta$ only. This conjecture can be verified in higher
orders also and is proved in the next section.

\section{Local deformations of the antibracket}

According to our conjecture,
let us look for the deformation of antibracket in the following form
\[
[ e^{pz},e^{qz}]_*=
F(\langle p,q\rangle ,\langle p,p\rangle ,\langle q,q\rangle ,N_{p},N_{q})e^{(p+q)z}.
\]
Then Jacobi identity acquires the form
\be
F(\langle p,q\rangle ,\langle p,p\rangle ,\langle q,q\rangle ,N_{p},N_{q})
F(\langle p+q,r\rangle ,\langle p+q,p+q\rangle ,\langle r,r\rangle ,N_{p+q},N_{r})
e^{(p+q+r)z}+ \mbox{cycle}=0
\label{ade1}
\ee

Because we are looking for deformations with even parameters only,
our conjecture gives
\begin{equation}
F(\langle p,q\rangle ,\langle p,p\rangle ,\langle q,q\rangle ,N_{p},N_{q})=\langle p,q\rangle +\langle p,p\rangle f(N_{p},N_{q})+\langle q,q\rangle f(N_{q}
,N_{p})\label{ade2}
\end{equation}
with some function $f$.

Substituting (\ref{ade2}) into (\ref{ade1}) and taking in account that
$
N_{p+q}=N_{p}+N_{q}
$
when acts on $e^{(p+q+r)z}$, we obtain the following system of equations
\begin{align}
&  f(N_{r},N_{p}+N_{q})
-    f(N_{r},N_{p}+1)
-    f(N_{r},N_{q}+1)=0
\label{adeu1}
\\
-&f(N_{p}+N_{q},N_{r})+
 f(N_{p}+1,N_{r})+
2   f(N_{p}+1,N_{q}+1)f(N_{p}+N_{q},N_{r})=0
\label{adeu2}
\\
&  f(N_{p},N_{q}+2)f(N_{p}+N_{q},N_{r})
-  f(N_{q},N_{r})f(N_{p},N_{q}+N_{r})
-
\nn
& \ \ \ \ \ \ -  f(N_{q},N_{p}+2)f(N_{p}+N_{q},N_{r})+
 f(N_{p},N_{r})f(N_{q},N_{r}+N_{p})=0
\label{adeu3}
\end{align}

The equation (\ref{adeu1}) gives
$
f(N_{r},N_{p}+N_{q})-f(N_{r},N_{p}+1)-f(N_{r},N_{q}+1)=0,
$
or
\begin{equation}
f(N_{r},N_{p})-f(N_{r},N_{p}+1)-f(N_{r},1)=0\label{adel}
\end{equation}

The equation (\ref{adel}) has an evident solution
\begin{equation}
f(N_{r},N_{p})=f(N_{r},0)(1-\frac{1}{2}N_{p})=\varphi(N_{r})(1-\frac{1}
{2}N_{p}),
\label{ader1}
\end{equation}
where $\varphi(N_{r})=f(N_{r},0).$

Let us substitute (\ref{ader1}) into (\ref{adeu2}).

We obtain
$
-f(N_{p}+N_{q},N_{r})+f(N_{p}+1,N_{r})+2f(N_{p}+1,N_{q}+1)f(N_{p}+N_{q}
,N_{r})=0,
$
or
\[
-\varphi(N_{p}+N_{q})(1-\frac{1}{2}N_{r})+\varphi(N_{p}+1)(1-\frac{1}{2}
N_{r})+2\varphi(N_{p}+1)(1-\frac{1}{2}(N_{q}+1))\varphi(N_{p}+N_{q}
)(1-\frac{1}{2}N_{r})=0,
\]
which implies
$
-\varphi(N_{p}+N_{q})+\varphi(N_{p}+1)+\varphi(N_{p}+1)(1-N_{q})\varphi
(N_{p}+N_{q})=0.
$

So
$
-\varphi(N_{q})+\varphi(1)+\varphi(1)(1-N_{q})\varphi(N_{q})=0,
$
and
$$
\varphi(N_{q})=\frac{a}{1+aN_{q}},
$$
where $a=\varphi(1)/(1-\varphi(1))$
.

Finally
\[
f(N_{p},N_{q})=\frac{a}{1+aN_{p}}(1-\frac{1}{2}N_{q}).
\]
This function satisfies the equation (\ref{adeu3}) as well.

So the deformation of antibracket is described by the formula
\be\label{pdefin}
[ e^{pz},e^{qz}]_*=\langle p,q\rangle +\langle p,p\rangle \frac{a}{1+aN_{p}}e^{pz}
(1-\frac{1}{2} N_{q})e^{qz}+
\langle q,q\rangle \frac{a}{1+aN_{a}}e^{qz}(1-\frac{1}{2}N_{p})e^{pz}.
\ee

Introduce the following 2-cochain
\be\label{k-cochain}
K_a(z|f,g)\stackrel  {def}=[f(z),g(z)]+
\left(\frac{a}{1+aN_z/2}\Delta f(z)\right)
\eE_z g(z)+
(-1)^{\varepsilon(f)}\left( \frac{a}{1+aN_{z}/2}\Delta g(z) \right )\eE_z f(z),
\ee
which is connected with (\ref{pdefin}) by Fourier transformation:
\be
K_a(z|f,g)=\int dp\tilde{f}(p)\int
dq\tilde{g}(q)(-1)^{\varepsilon(g)}K_a(z|e^{ipz},e^{iqz}),
\ee
where
\be
f(z)=\int dp\tilde{f}(p)e^{ipz}.
\ee

For Jacobiator
$J_{K_a}(z|f,g,h)\stackrel {def}=
K_a(z|f,K_a(\cdot |g,h))+\mbox{cycle}(f,g,h)$,
analogous relation take place:
\[
J_{K_a}(z|f,g,h)=-\int dp\tilde{f}(p)\int dq\tilde{g}(q) \int
dr\tilde{h}(r)(-1)^{\varepsilon(f)\varepsilon(h)+
\varepsilon(f)+\varepsilon(g)+\varepsilon(h)}J_{K_a}(z|e^{ipz},e^{iqz},e^{irz}).
\]
So, because $J_{K_a}(z|e^{ipz},e^{iqz},e^{irz})=0$ by construction, we have that
$J_{K_a}(z|f,g,h)=0$ for all $f,g,h\in \mathbf D_n$.

\section{The proof of theorem \ref{thdef}}

Let $c_i\in\K$ be some sequence. Define $c_{[k]}$ as
$c_{[k]}=\sum_{i=1}^k \hbar^i c_i$.
Consider once again the equation (\ref{1.1a}) for decomposition (\ref{decc}).

The first order in $\hbar$ gives
$$C(z|f,g)=[f(z),g(z)]+
c_1\hbar m(z|f,g) + \hbar d_{1}^{\mathrm{ad}}M_{1D}(z|f,g)+O(\hbar^2).$$
Let us make similarity transformation with the operator $T_1$ of the form
$T_1(z|f)=f(z)-\hbar M_{1D}(z|f)+O(\hbar^2)$, and keep the notation $C(z|f,g)$
for resulting form. It can be presented as
\[
C(z|f,g)=K_{c_{[1]}}(z|f,g)+\hbar^2C_2 (z|f,g)+O(\hbar^3).
\]
Then, we have from Eq. (\ref{1.1a})
\begin{eqnarray*}
d_{2}^{\mathrm{ad}}C_{2}(z|f,g,h)=0,\;\varepsilon
_{C_{21}}=1\;\Longrightarrow
\end{eqnarray*}
\begin{equation*}
C_{2}(z|f,g)=c_{2}m(z|f,g)+d_{1}^{\mathrm{ad}}M_{2D}(z|f,g),
\end{equation*}
The condition $C_{2}(z|f,g)\in D$ gives $M_{2D}(z|f)\in D$.
Making similarity transformation with the operator $T_2$ of the form
$T_2(z|f)=f(z)-\hbar M_{2D}(z|f)+O(\hbar^3)$, and present the resulting 2-cochain,
which we denote once again as $C(z|f,g)$, in a form
\[
C(z|f,g)=K_{c_{[2]}}(z|f,g)+\hbar^3C_3 (z|f,g)+O(\hbar^4).
\]
Then, we have from Eq. (\ref{1.1a})
\begin{eqnarray*}
d_{2}^{\mathrm{ad}}C_{3}(z|f,g,h)=0,
\end{eqnarray*}
and so on.

Finally, we obtain, that general solution of Eq. (\ref{1.1a}) has
the form up to similarity transformation
\[
[f(z),g(z)]_*=K_{c_{[\infty]}}(z|f,g),
\]
that finishes the proof of the theorem \ref{thdef}.

\section{Central extension of antiPoisson superalgebra}

First, we will consider some notation and general formulas.

Let  $L=\mathbf D_1$.
According to theorem \ref{thtr} it has the central
extension
$\bbreve L=L\oplus c$. If $f\in L$, then $\bbreve{f}=f+a_{i}c_{i}\in \bbreve{L}$,
$a_{i}c_{i}\in c$, $i=1,2$, $\varepsilon (c_{1})=1$, $\varepsilon (c_{2})=0$ (
$\varepsilon (c_{i})=\varepsilon _{i}=i$), $a_i\in \K$. The antibracket in $L$
will be denoted $[f,g]=-(-1)^{\epsilon (f)\epsilon (g)}[g,f]$.
The bracket in $\bbreve{L}$ will be denoted $[\bbreve{f},\bbreve{g}]_{\mathrm{ce
}}=-(-1)^{\epsilon (\bbreve{f})\epsilon (\bbreve{g})}[\bbreve{g},
\bbreve{f}]_{\mathrm{ce}}$,
\[
[ f,g]_{\mathrm{ce}}=[f,g]+\mu _{i}(f,g)c_{i},\;[\bbreve{f},c_{i}]_{
\mathrm{ce}}=0,
\]
where $\mu _{i}(f,g)=-(-1)^{\epsilon (f)\epsilon (g)}\mu
_{i}(g,f)$ are 2-cocycles in the trivial
representation of the algebra $L$ defined by Eq. (\ref{mu}) in
theorem \ref{thtr}.

We will consider 1-cochains and 2-cochains on $\bbreve{L}$ with coefficients
in adjoint representation:
\begin{eqnarray*}
\bbreve{M}_{1}(\bbreve{f}) &=&M_{1}(z|\bbreve{f})+m_{1i}(\bbreve{f})c_{i}\in
\bbreve{L},\;M_{1}(z|\bbreve{f})\in L,\;m_{1i}(\bbreve{f})\in \K,\; \\
\varepsilon (\bbreve{M}_{1}) &=&\varepsilon (M_{1})=\varepsilon
(m_{12})=0,\;\varepsilon (m_{11})=1\;(\varepsilon _{m_{1i}}\equiv
\varepsilon _{i}=i);
\end{eqnarray*}
\begin{eqnarray*}
\bbreve{M}_{2}(\bbreve{f},\bbreve{g}) &=&M_{2}(z|\bbreve{f},\bbreve{g})+m_{2i}(
\bbreve{f},\bbreve{g})c_{i}\in \bbreve{L},\;M_{2}(z|\bbreve{f},\bbreve{g})\in
L,\;m_{2i}(\bbreve{f},\bbreve{g})\in \K, \\
\bbreve{M}_{2}(\bbreve{g},\bbreve{f}) &=&-(-1)^{\epsilon (\bbreve{f}
)\epsilon (\bbreve{g})}\bbreve{M}_{2}(\bbreve{f},\bbreve{g}),\;\bbreve{M
}_{2}(c_{1},c_{1})=0, \\
\varepsilon (\bbreve{M}_{2}) &=&\varepsilon (M_{2})=\varepsilon
(m_{22})=1,\;\varepsilon (m_{21})=0\;(\varepsilon _{m_{2i}}\equiv
\varepsilon _{i}+1=i+1).
\end{eqnarray*}

The differential $\bbreve{d}^{\mathrm{ad}}$
is defined on these (and on all others) forms:
\begin{equation}
\bbreve{d}_{1}^{\mathrm{ad}}\bbreve{M}_{1}(\bbreve{f},\bbreve{g})=[\bbreve{M}_{1}(
\bbreve{f}),\bbreve{g}]_{\mathrm{ce}}-(-1)^{\epsilon (\bbreve{f}
)\epsilon (\bbreve{g})}[\bbreve{M}_{1}(\bbreve{g}),\bbreve{f}]_{
\mathrm{ce}}-\bbreve{M}_{1}([\bbreve{f},\bbreve{g}]_{\mathrm{ce}
})\;\Longrightarrow  \label{c.1.0}
\end{equation}
\begin{eqnarray}
&&\,\bbreve{d}_{1}^{\mathrm{ad}}\bbreve{M}_{1}(f,g)=d_{1}^{\mathrm{ad}
}M_{1}(z|f,g)-\mu _{i}(f,g)M_{1}(z|c_{i})+\gamma _{i}(f,g)c_{i},  \nonumber
\\
&&\;\gamma _{i}(f,g)=\mu _{i}(M_{1}(|f),g)-(-1)^{\epsilon
(f)\epsilon (g)}\mu _{i}(M_{1}(|g),f)+  \nonumber \\
&&+d_{1}^{\mathrm{tr}}m_{1i}(f,g)-(-1)^{i(j+1)}\mu _{j}(f,g)m_{1i}(c_{j}),
\label{c.1.1} \\
&&\bbreve{d}_{1}^{\mathrm{ad}}\bbreve{M}_{1}(c_{i},g)=[M_{1}(z|c_{i}),g(z)]+
\mu _{j}(M_{1}(|c_{i}),g)c_{j},  \label{c.1.2} \\
&&\bbreve{d}_{1}^{\mathrm{ad}}\bbreve{M}_{1}(c_{i},c_{j})\equiv 0,  \nonumber
\\
&&d_{1}^{\mathrm{tr}}m_{1i}(f,g)=-m_{1i}([f,g]).  \nonumber
\end{eqnarray}

\begin{eqnarray}
\bbreve{d}_{2}^{\mathrm{ad}}\bbreve{M}_{2}(\bbreve{f},\bbreve{g},\bbreve{h}
)&=&-(-1)^{\epsilon (\bbreve{f})\epsilon (\bbreve{h}
)}\{(-1)^{\epsilon (\bbreve{f})\epsilon (\bbreve{h})}[\bbreve{M
}_{2}(\bbreve{f},\bbreve{g}),\bbreve{h}]_{\mathrm{ce}}+ \nn
&&\,+(-1)^{\epsilon (\bbreve{f})\epsilon (\bbreve{h})}\bbreve{M}
_{2}([\bbreve{f},\bbreve{g}]_{\mathrm{ce}},\bbreve{h})+\mathrm{cycle}(f,g,h)\}=
\nn
\,&=&-(-1)^{\epsilon (\bbreve{f})\epsilon (\bbreve{h}
)}\{(-1)^{\epsilon (\bbreve{f})\epsilon (\bbreve{h})}[M_{2}(z|
\bbreve{f},\bbreve{g}),\bbreve{h}]_{\mathrm{ce}}+ \nn
&&+(-1)^{\epsilon (\bbreve{f})\epsilon (\bbreve{h})}M_{2}(z|[
\bbreve{f},\bbreve{g}]_{\mathrm{ce}},\bbreve{h})+ \nn
&&\,+(-1)^{\epsilon (\bbreve{f})\epsilon (\bbreve{h})}m_{2i}([
\bbreve{f},\bbreve{g}]_{\mathrm{ce}},\bbreve{h})c_{i}+\mathrm{cycle}
(f,g,h)\}\;\Longrightarrow
\end{eqnarray}
\begin{eqnarray}
&&\bbreve{d}_{2}^{\mathrm{ad}}\bbreve{M}_{2}(f,g,h)=d_{2}^{\mathrm{ad}
}M_{2}(z|f,g,h)+\{-\mu _{i}(M_{2}(|f,g),h)+  \nonumber \\
&&+(-1)^{\epsilon (g)\epsilon (h)}\mu
_{i}(M_{2}(|f,h),g)-(-1)^{\epsilon (f)(\epsilon (g)+\epsilon
(h))}\mu _{i}(M_{2}(|g,h),f)-  \nonumber \\
&&-m_{2i}(\mu _{j}(f,g)c_{j},h)+(-1)^{\epsilon (g)\epsilon
(h)}m_{2i}(\mu _{j}(f,h)c_{j},g)-  \nonumber \\
&&-(-1)^{\epsilon (f)(\epsilon (g)+\epsilon (h))}m_{2i}(\mu
_{j}(g,h)c_{j},g)+d_{2}^{\mathrm{tr}}m_{2i}(f,g,h)\}c_{i}-  \nonumber \\
&&-M_{2}(z|\mu _{j}(|f,g)c_{j},h)+(-1)^{\epsilon (g)\epsilon
(h)}M_{2}(z|\mu _{j}(|f,h)c_{j},g)-  \nonumber \\
&&-(-1)^{\epsilon (f)(\epsilon (g)+\epsilon (h))}M_{2}(z|\mu
_{j}(|g,h)c_{j},f),  \label{c.2.1} \\
&&d_{2}^{\mathrm{tr}}m_{2i}(f,g,h)=-\{m_{2i}([f,g],h)-(-1)^{\epsilon
(g)\epsilon (h)}m_{2i}([f,h],g)-m_{2i}(f,[g,h])\}  \nonumber
\end{eqnarray}
\begin{eqnarray}
&&\bbreve{d}_{2}^{\mathrm{ad}}\bbreve{M}_{2}(c_{i},f,g)=-d_{1}^{\mathrm{ad}}
\tilde{M}_{1i}(z|f,g)+M_{2}(z|c_{i},\mu _{l}(f,g)c_{l})+  \nonumber \\
&&+\{-\mu _{j}(\tilde{M}_{1i}(|f),g)+(-1)^{\epsilon (f)\varepsilon
(g)}\mu _{j}(\tilde{M}_{1i}(|g),f)+  \nonumber \\
&&+m_{2j}(c_{i},[f,g])+m_{2j}(c_{i},\mu _{l}(f,g)c_{l})\}c_{j},
\label{c.2.2} \\
&&\tilde{M}_{1i}(z|f)=M_{2}(z|c_{i},f),  \nonumber
\end{eqnarray}
\begin{equation}
\bbreve{d}_{2}^{\mathrm{ad}}\bbreve{M}
_{2}(c_{i},c_{j},f)=-[M_{2}(z|c_{i},c_{j}),f(z)]-\mu
_{k}(M_{2}(|c_{i},c_{j}),f)c_{k},  \label{c.2.3}
\end{equation}
\[
\bbreve{d}_{2}^{\mathrm{ad}}\bbreve{M}_{2}(c_{i},c_{j},c_{k})\equiv 0.
\]

\section{Deformation of central extension of antiPoisson superalgebra}

\subsection{Second central adjoint cohomology}

The cohomology equation is
\[
\bbreve{d}_{2}^{\mathrm{ad}}\bbreve{M}_{2}(f,g,h)=0.
\]

It follows from Eq. (\ref{c.2.3})
\[
\frac{\partial }{\partial z^{A}}M_{2}(z|c_{i},c_{j})=0\;\Longrightarrow
\;M_{2}(z|c_{i},c_{j})=a_{2|ij},\;a_{2|11}=0,\;a_{2|12}=-a_{2|21}.
\]
Eq. (\ref{c.2.2}) takes the form
\begin{eqnarray}
&&\,[\tilde{M}_{1i}(z|f),g(z)]-(-1)^{\epsilon (f)\epsilon (g)}[\tilde{M}
_{1i}(z|g),f(z)]-\tilde{M}_{1i}(z|[f,g])=a_{2|ij}\mu _{j}(f,g),
\label{c.3.1} \\
&&m_{2j}(c_{i},[f,g])+m_{2j}(c_{i},c_{l})\mu _{l}(f,g)=\mu _{j}(\tilde{M}
_{1i}(|f),g)-(-1)^{\epsilon (f)\epsilon (g)}\mu _{j}(\tilde{M}_{1i}(|g),f).
\label{c.3.2}
\end{eqnarray}

First, consider Eq. (\ref{c.3.1}). Let
\[
z\cap \mathrm{supp}(f)=z\cap \mathrm{supp}(g)=\varnothing .
\]
We obtain from Eq. (\ref{c.3.1})
\[
\tilde{M}_{1i}(z|[f,g])=-a_{2|ij}\mu _{j}(f,g).
\]
Let $g=x$ or $g=\xi $ for $\{z\}\subset \mathrm{supp}(f)$. Then in both these
cases $\mu _{j}(f,g)=0$ and we find $\tilde{M}_{1i}(z|\partial _{A}f)=0$
(for $z\cap \mathrm{supp}(f)=\varnothing $). Let $g=x\xi $. Then $\mu
_{j}(f,g)=0$ and we obtain
\[
\tilde{M}_{1i}(z|f)=0\;\mathrm{for}\;z\cap \mathrm{supp}(f)=\varnothing
\;\Longrightarrow \;a_{2|ij}=0\;\Longrightarrow
\]
\[
\tilde{M}_{1i}(z|f)=t_{i}^{0}\mathcal{E}_{z}f(z)+t_{i}\Delta
f(z)+[t_{1i}(z),f(z)].
\]
The summand $[t_{1i}(z),f(z)]$ can be included in exact form $\bbreve{d}_{1}^{
\mathrm{ad}}\bbreve{M}_{1}(f,g)$ with $M_{1}(z|f)=m_{1i}(\bbreve{f})=0$, $
M_{1}(z|c_{i})=t_{1i}(z)$, such that we have for the form $\bbreve{M}
_{2}^{\prime }(f,g,h)=\bbreve{M}_{2}(f,g,h)-\bbreve{d}_{1}^{\mathrm{ad}}\bbreve{
M}_{1}(f,g)$ (omitting primes)
\begin{equation}
m_{2j}(c_{i},[f,g])+m_{2j}(c_{i},c_{l})\mu _{l}(f,g)=t_{i}^{0}\mu _{j}(
\mathcal{E}_{z}f,g)+t_{i}\mu _{j}(\Delta f,g)-(-1)^{\epsilon (f)\epsilon
(g)}(f\leftrightarrow g).  \label{c.3.3}
\end{equation}
Choosing $g=x,\xi ,x\xi $, we obtain, in the same way as above,
\[
m_{2j}(c_{i},f)=0.
\]

Eq. (\ref{c.3.3}) takes the form (taking in account that $\mu _{2}(\Delta
f,g)=\mu _{1}(f,g)$)
\begin{eqnarray}
&&0=t_{1}^{0}[\mu _{1}(\mathcal{E}_{z}f,g)-(-1)^{\epsilon (f)\epsilon
(g)}(f\leftrightarrow g)],  \label{c.3.4a} \\
&&m_{21}(c_{2},c_{1})\mu _{1}(f,g)=t_{2}^{0}[\mu _{1}(\mathcal{E}
_{z}f,g)-(-1)^{\epsilon (f)\epsilon (g)}(f\leftrightarrow g)],
\label{c.3.4b} \\
&&m_{21}(c_{i},c_{2})\mu _{2}(f,g)=0,  \label{c.3.4c} \\
&&0=2t_{1}\mu _{1}(f,g),  \label{c.3.4d} \\
&&m_{22}(c_{2},c_{1})\mu _{1}(f,g)=2t_{2}\mu _{1}(f,g),  \label{c.3.4e} \\
&&m_{22}(c_{i},c_{2})\mu _{2}(f,g)=t_{i}^{0}[\mu _{2}(\mathcal{E}
_{z}f,g)-(-1)^{\epsilon (f)\epsilon (g)}(f\leftrightarrow g)].
\label{c.3.4f}
\end{eqnarray}
It follows from (\ref{c.3.4c}) that $m_{21}(c_{i},c_{j})=0$. Then Eqs. (\ref
{c.3.4a}) and (\ref{c.3.4b}) give $t_{i}^{0}=0$. Then it follows from
Eq. (\ref{c.3.4f}) that $m_{22}(c_{i},c_{j})=0$ and finally Eqs. (\ref
{c.3.4d}) and (\ref{c.3.4e}) give $t_{i}=0$. Thus, we obtained (up to
coboundary)
\[
M_{2}(z|c_{i},f)=m_{2i}(c_{j},f)=M_{2}(z|c_{i},c_{j})=m_{2i}(c_{j},c_{l})=0.
\]

Eq. (\ref{c.2.1}) takes the form
\begin{eqnarray}
&&d_{2}^{\mathrm{ad}}M_{2}(z|f,g,h)=0,  \label{c.3.5a} \\
&&d_{2}^{\mathrm{tr}}m_{2i}(f,g,h)=\mu _{i}(M_{2}(|f,g),h)-(-1)^{\epsilon
(g)\epsilon (h)}\mu _{i}(M_{2}(|f,h),g)+  \nonumber \\
&&\,+(-1)^{\epsilon (f)(\epsilon (g)+\epsilon (h))}\mu _{i}(M_{2}(|g,h),f).
\label{c.3.5b}
\end{eqnarray}
Firstly, consider Eq. (\ref{c.3.5a}). Under condition $\varepsilon (M_{2})=1$,
its general solution has the form
\begin{eqnarray*}
M_{2}(z|f,g) &=&am(z|f,g)+d_{1}^{\mathrm{ad}}\sigma _{1}(z|f,g).
\end{eqnarray*}
The summand $d_{1}^{\mathrm{ad}}\sigma _{1}(z|f,g)$ can be included in coboundary
$\bbreve{d}_{1}^{\mathrm{ad}}\bbreve{M}_{1}(f,g)$ (additionally to $
M_{2}(z|f,g)$, only $m_{2i}(f,g)$ is modified) with $M_{1}(z|f)=\sigma
_{1}(z|f)$, $m_{1i}(\bbreve{f})=M_{1}(z|c_{i})=0$, such that we have for the
form $\bbreve{M}_{2}^{\prime }(f,g,h)=\bbreve{M}_{2}(f,g,h)-\bbreve{d}_{1}^{
\mathrm{ad}}\bbreve{M}_{1}(f,g)$ (omitting primes)
\begin{eqnarray}
&&M_{2}(z|f,g)=am(z|f,g),  \nonumber \\
&&d_{2}^{\mathrm{tr}}m_{2i}(f,g,h)=(-1)^{\epsilon (f)\epsilon
(h)}aJ_{i}(f,g,h),  \label{c.3.6} \\
J_{i}(f,g,h) &=&(-1)^{\epsilon (f)\epsilon (h)}\mu _{i}(m(|f,g),h)+
\mathrm{cycle}(f,g,h).  \nonumber
\end{eqnarray}
Consider Eq. (\ref{c.3.6}) in for $i=2$,
\begin{eqnarray}
&&d_{2}^{\mathrm{tr}}m_{22}(f,g,h)=(-1)^{\epsilon (f)\epsilon
(h)}aJ_{2}(f,g,h),  \label{c.3.7} \\
&&J_{2}(f,g,h)=(-1)^{\epsilon (f)\epsilon (h)+\epsilon (g)}\int dz[\Delta
f(z)N_{z}g(z)+  \nonumber \\
&&+(-1)^{\varepsilon (f)}N_{z}f(z)\Delta g(z)]\partial _{x}^{2}h(z)+\mathrm{
cycle}(f,g,h).  \nonumber
\end{eqnarray}

Let
\[
\mathrm{supp}(h)\cap \left[ \mathrm{supp}(f)\cup \mathrm{supp}(g)\right]
=\varnothing .
\]
Then, r.h.s. of Eq. (\ref{c.3.7}) equals zero and we have
\begin{eqnarray*}
&&m_{22}(f,g)=\sum_{k=0}^{K}\int dzF_{1}^{k}(z)(-1)^{\varepsilon (f)}\left(
[\partial _{x}^{2k}f(z)]g(z)+f(z)\partial _{x}^{2k}g(z)\right) + \\
&&+\sum_{k=1}^{K}\int dzF_{2}^{k}(z)\left( (-1)^{\varepsilon (f)}[\partial
_{x}^{2k-1}\partial _{\xi }f(z)]g(z)+f(z)\partial _{x}^{2k-1}\partial _{\xi
}g(z)\right) .
\end{eqnarray*}
Take the functions in the form $f(z)=e^{zp}$, $g(z)=e^{zq}$, $\
h(z)\rightarrow e^{-z(p+q)}h(z)$, and consider the terms of the highest
order in $p$ and $q$ which equals $2K+2$ for l.h.s. of Eq. (\ref{c.3.7})
and $5$ for r.h.s. of Eq. (\ref{c.3.7}). Thus,
we have
\begin{eqnarray*}
&&F_{1}^{k}(z)=0,\;k\geq 2,\;F_{2}^{k}(z)=0,\;k\geq 3, \\
&&m_{22}(f,g)=\sum_{k=0}^{1}\int dzF_{1}^{k}(z)(-1)^{\varepsilon (f)}\left(
[\partial _{x}^{2k}f(z)]g(z)+f(z)\partial _{x}^{2k}g(z)\right) + \\
&&+\sum_{k=1}^{2}\int dzF_{2}^{k}(z)[(-1)^{\varepsilon (f)}(\partial
_{x}^{2k-1}\partial _{\xi }f(z))g(z)+f(z)\partial _{x}^{2k-1}\partial _{\xi
}g(z)].
\end{eqnarray*}
Consider in Eq. (\ref{c.3.7}) the summands which are of the sixth and fifth
orders in $p$, $q$. Only $F_{2}^{2}(z)$ takes part in l.h.s. of Eq. (\ref
{c.3.7}). The sixth order in $p$, $q$ terms are canceled identically and we
obtain for the fifth order terms
\begin{eqnarray*}
&&\,-\partial
_{x}F_{2}^{2}(z)(2p_{0}^{3}+3p_{0}^{2}q_{0}-3p_{0}q_{0}^{2}-2q_{0}^{3})p_{1}q_{1}-
\\
&&\,-F_{2}^{2}(z)\overleftarrow{\partial }_{\xi
}[(2p_{0}^{3}q_{0}+3p_{0}^{2}q_{0}^{2}+p_{0}q_{0}^{3})p_{1}+(p_{0}^{3}q_{0}+3p_{0}^{2}q_{0}^{2}+2p_{0}q_{0}^{3})q_{1}]=
\\
&&
\,=ax[(2p_{0}^{3}q_{0}+3p_{0}^{2}q_{0}^{2}+p_{0}q_{0}^{3})p_{1}+(p_{0}^{3}q_{0}+3p_{0}^{2}q_{0}^{2}+2p_{0}q_{0}^{3})q_{1}]-
\\
&&-a\xi (2p_{0}^{3}+3p_{0}^{2}q_{0}-3p_{0}q_{0}^{2}-2q_{0}^{3})p_{1}q_{1}.
\end{eqnarray*}
where $p=(p_{0},p_{1})$, $q=(q_{o},q_{1})$, $\varepsilon (p_{0})=\varepsilon
(q_{0})=0$, $\varepsilon (p_{1})=\varepsilon (q_{1})=1$. Setting $q_{0}=\pm
p_{0}$, we find
\[
F_{2}^{2}(z)\overleftarrow{\partial }_{\xi }=-ax,\;\partial
_{x}F_{2}^{2}(z)=a\xi \;\Longrightarrow
\]
\[
a=0.
\]

It follows from $$m_{22}(f,g)=\int dz F_{2}^{2}(z)[(-1)^{\varepsilon
(f)}(\partial _{x}^{3}\partial _{\xi }f(z))g(z)+f(z)\partial
_{x}^{3}\partial _{\xi }g(z)], \quad \varepsilon (m_{22})=1$$ that $\varepsilon
(F_{2}^{2}(z))=1$ and so $F_{2}^{2}(z)=\xi F(x)$.
So
$$m_{22}(f,g)=\int dxF(x)[f_{0}(x)\partial _{x}^{3}g_{1}(z)-\partial
_{x}^{3}f_{1}(z))g_{0}(z)].$$
 It is easy to see that $d_{2}^{\mathrm{tr}
}m_{22}(f_{0},g_{0},h_{0})=$ $d_{2}^{\mathrm{tr}
}m_{22}(f_{0},g_{0},h_{1})=d_{2}^{\mathrm{tr}}m_{22}(f_{1},g_{1},h_{1})=0$,
and
\[
d_{2}^{\mathrm{tr}}m_{22}(f_{0},g_{1},h_{1})=\int dxf_{0}[F(\partial
_{x}^{3}g_{1}\partial _{x}h_{1}-\partial _{x}g_{1}\partial
_{x}^{3}h_{1})-\partial _{x}F(\partial _{x}^{3}g_{1}h_{1}-g_{1}\partial
_{x}^{3}h_{1})],
\]
where $\varphi (z)\equiv \varphi _{0}(x)+\xi \varphi _{1}(x)$ for any
function $\varphi (z)$. Analogously, we find that $
J_{2}(f_{0},g_{0},h_{0})=J_{2}(f_{0},g_{0},h_{1})=$ $
J_{2}(f_{1},g_{1},h_{1})=0$ and
\begin{eqnarray*}
&&J_{2}(f_{0},g_{1},h_{1})=\int dxf_{0}[x(\partial _{x}^{3}g_{1}\partial
_{x}h_{1}-\partial _{x}g_{1}\partial _{x}^{3}h_{1})+(\partial
_{x}^{3}g_{1}h_{1}+2\partial _{x}^{2}g_{1}\partial _{x}h_{1}- \\
&&-2\partial _{x}g_{1}\partial _{x}^{2}h_{1}-g_{1}\partial _{x}^{3}h_{1})].
\end{eqnarray*}
It follows from Eq. (\ref{c.3.7})
\begin{eqnarray*}
&&F(\partial _{x}^{3}g_{1}\partial _{x}h_{1}-\partial _{x}g_{1}\partial
_{x}^{3}h_{1})-\partial _{x}F(\partial _{x}^{3}g_{1}h_{1}-g_{1}\partial
_{x}^{3}h_{1})= \\
&&\,=a[x(\partial _{x}^{3}g_{1}\partial _{x}h_{1}-\partial _{x}g_{1}\partial
_{x}^{3}h_{1})+ \\
&&\,+(\partial _{x}^{3}g_{1}h_{1}+2\partial _{x}^{2}g_{1}\partial
_{x}h_{1}-2\partial _{x}g_{1}\partial _{x}^{2}h_{1}-g_{1}\partial
_{x}^{3}h_{1})]\;\Longrightarrow
\end{eqnarray*}
\[
a=F(x)=F_{2}^{2}(x)=0.
\]

Thus, we have
\[
d_{2}^{\mathrm{tr}}m_{2i}(f,g,h)=0\;\Longrightarrow
\]
\[
m_{2i}(f,g)=a_{ij}\mu _{j}(f,g)+d_{1}^{\mathrm{tr}}\nu _{i}(f,g).
\]
This expression corresponds to exact form $\bbreve{d}_{1}^{\mathrm{ad}}\bbreve{
M}_{1}(\bbreve{f},\bbreve{g})$ with $M_{1}(\bbreve{f})=0$, $m_{1i}(f)$, $
m_{1i}(c_{j})=-(-1)^{i(j+1)}a_{ij}$. Finally, we have obtained that $H_{
\mathrm{ad}}^{2}=0$ in subspace of 2-forms $\bbreve{M}_{1}(\bbreve{f},\bbreve{g}
)$ with $\ \varepsilon (\bbreve{M}_{1})=1$. By the same way as above, we can
prove that
\[
\bbreve{d}_{1}^{\mathrm{ad}}\bbreve{M}_{1}(\bbreve{f},\bbreve{g})\in
\bbreve L\;\Longrightarrow \;\bbreve{M}_{1}(\bbreve{f})\in \bbreve L,
\]
such that the central extension of  antiPoisson algebra has no nontrivial
deformations.

\end{document}